\documentclass[12pt,usenames,dvipsnames]{article}

\usepackage{latexsym}
\usepackage{amssymb,amsfonts,amsmath}
\usepackage{graphicx} 
\usepackage{indentfirst}
\usepackage{bbm}
\usepackage{amssymb}
\usepackage{verbatim}
\usepackage{amsmath, amsthm,amssymb}
\usepackage{mathrsfs}
\usepackage{hyperref}
\usepackage{amsfonts}
\usepackage{dsfont}
\usepackage{cite}
\usepackage{xcolor}
\usepackage{enumerate}
\parskip=\medskipamount

\newcommand {\cE}{{\cal E}}

\newcommand {\cN}{{\cal N}}

\def\a{\alpha}

\def\b{\beta}

\def\d{\delta}
\def\e{\epsilon}

\def\g{\gamma}
\def\G{\Gamma}

\def\j{\psi}
\def\k{\kappa}
\def\l{\lambda}
\def\m{\mu}

\def\o{\omega}

\def\q{\theta}
\def\r{\rho}
\def\s{\sigma}
\def\t{\tau}

\def\z{\zeta}

\def\J{\Psi}

\def\O{\Omega}

\def\U{\Upsilon}

\def\rd{{\rm d}}
\def\ri{{\rm i}}

\newcommand{\ad}{{\dot{\alpha}}}                           %new
\newcommand{\bd}{{\dot{\beta}}}                            %new
\newcommand{\ve}{\varepsilon}                            %new
\renewcommand{\aa}{{\a\ad}}
\newcommand{\bb}{{\b\bd}}
\newcommand{\pa}{\partial}                           %new
\newcommand{\hf}{\frac12}
\newcommand{\vf}{\varphi}
\newcommand{\be}{\begin{equation}}
\newcommand{\ee}{\end{equation}}
\newcommand{\bea}{\begin{eqnarray}}
\newcommand{\eea}{\end{eqnarray}}
\newcommand{\non}{\nonumber}
\newcommand{\bm}[1]{\mbox{\boldmath$#1$}}

\def\double #1{#1{\hbox{\kern-2pt $#1$}}}

\newcommand{\gd}{{\dot\g}}
\newcommand{\dd}{{\dot\d}}

\newcommand{\Nabla}{\bm{\nabla}}
\newcommand{\bNabla}{\bar{\bm{\nabla}}}

\newif\ifdtup

\newcommand{\bsubeq}{\begin{subequations}}
\newcommand{\esubeq}{\end{subequations}}
\numberwithin{equation}{section}

\newcommand{\sSU}{\mathsf{SU}}

\newcommand{\sU}{\mathsf{U}}

\begin{document}

\begin{titlepage}
\begin{flushright}
November, 2020 \\
\end{flushright}
\vspace{5mm}

\begin{center}
{\Large \bf 

Generalised superconformal higher-spin multiplets

}
\end{center}

\begin{center}

{\bf Sergei M. Kuzenko, Michael Ponds and Emmanouil S. N. Raptakis} \\
\vspace{5mm}

\footnotesize{
{\it Department of Physics M013, The University of Western Australia\\
35 Stirling Highway, Perth W.A. 6009, Australia}}  
~\\
\vspace{2mm}
~\\
Email: \texttt{ 
sergei.kuzenko@uwa.edu.au, michael.ponds@research.uwa.edu.au, emmanouil.raptakis@research.uwa.edu.au}\\
\vspace{2mm}

\end{center}

\begin{abstract}
\baselineskip=14pt

We propose generalised $\mathcal{N}=1$ superconformal higher-spin (SCHS)  gauge multiplets
of depth $t$, $\Upsilon_{\a(n)\ad(m)}^{(t)}$, with $n\geq m \geq 1$. 
At the component level, for $t>2$ they contain generalised conformal higher-spin (CHS) gauge fields with depths $t-1$, $t$ and $t+1$. The supermultiplets with $t=1$ and $t=2$ include both ordinary and generalised CHS gauge fields.
Super-Weyl and gauge invariant actions describing the dynamics of $\U_{\a(n)\ad(m)}^{(t)}$ on conformally-flat superspace backgrounds are then derived. For the  case $n=m=t=1$, corresponding to the maximal-depth conformal graviton supermultiplet, we extend this action to Bach-flat backgrounds. Models for superconformal non-gauge multiplets, which are expected to play an important role in the Bach-flat completions of the models for $\U^{(t)}_{\a(n)\ad(m)}$, are also provided.  Finally we show that, on Bach-flat backgrounds, requiring gauge and Weyl invariance does not always determine a model for a CHS field uniquely.

\end{abstract}
\vspace{5mm}

\vfill

\vfill
\end{titlepage}

\newpage
\renewcommand{\thefootnote}{\arabic{footnote}}
\setcounter{footnote}{0}

\tableofcontents{}
\vspace{1cm}
\bigskip\hrule

\allowdisplaybreaks

%\pagebreak

\section{Introduction}

Ever since its inception in 1985 \cite{FT}, conformal higher-spin (CHS) theory has been the recipient of sustained interest from the physics community. There are a number of reasons for this,
including a consistent Lagrangian formulation of bosonic CHS fields at the cubic \cite{FL,FL2} and full non-linear level \cite{Tseytlin, Segal} (see \cite{ BJM1, BJM2, Bonezzi} for later developments).
One of its central open problems is the construction of gauge invariant models for CHS fields on curved gravitational backgrounds, where much effort has been directed \cite{NT, GrigorievT,KMT, BeccariaT, Manvelyan, KP19, KP19-2, KPR}. As is well known, consistent models for conformal spin-$3/2$ and spin-$2$ fields may be formulated at most on Bach-flat backgrounds.
While this is thought also to be true for spins greater than two, recent studies \cite{GrigorievT, BeccariaT, KP19-2, KPR} indicate that in this case it is necessary to switch on non-minimal couplings to subsidiary conformal fields. 

Given positive integers $n\geq m \geq 1$, an ordinary conformal gauge field $h_{\a(n)\ad(m)} (x)$ on curved space is characterised by the gauge transformation
\begin{align}
\delta_{\ell}h_{\a(n)\ad(m)}=\nabla_{(\a_1(\ad_1}\ell_{\a_{2}\dots\a_{n})\ad_{2}\dots\ad_m)}~,
\end{align}
and carries conformal weight $\big(2-\frac{1}{2}(n+m)\big)$. Here $\nabla_{\a\ad}$ is the conformally covariant derivative \eqref{10.1}. Due to their low conformal weights and consequently high derivative Lagrangians, CHS fields are notoriously difficult to work with. This is why no closed form models for spin $s>2$ have been constructed on Bach-flat backgrounds. However, there exists a broader class of conformal gauge fields, the so-called generalised ones $h^{(t)}_{\a(n)\ad(m)}$. The latter carry the conformal weight $\big( t+1-\frac{1}{2}(n+m)\big)$ and are characterised by gauge transformations with depth $t$
\begin{align}
\delta_{\ell}h^{(t)}_{\a(n)\ad(m)}=\nabla_{(\a_1(\ad_1}\cdots\nabla_{\a_t\ad_t}\ell^{(t)}_{\a_{t+1}\dots\a_{n})\ad_{t+1}\dots\ad_m)}~,\qquad 1\leq t \leq \text{min}(n,m)~. \label{1}
\end{align}
 By virtue of their relatively high conformal weights and consequently lower derivative Lagrangians, they can provide a much friendlier environment with which to study CHS theory in. Indeed, gauge invariant actions on Bach-flat backgrounds have been explicitly derived for conformal spin $s=5/2$ and $s=3$ gauge fields with maximal depth \cite{KP19-2}.

 Historically, the first generalised conformal field appeared in the seminal work by Deser and Nepomechie \cite{DeserN1, DeserN2}, where they discussed the maximal-depth conformal graviton (corresponding to the case with $n=m=t=2$). 
 The concept was later extended to tensors of a generic symmetry type by Vasiliev in \cite{Vasiliev}, where the corresponding conformal and gauge invariant actions in $d$-dimensional Minkowski space $\mathbb{M}^d$  were also given (see \cite{BG,Barnich,GrigorievH,Skvortsov,Mkrtchyan} for more recent related studies). In the cases $d=3,4,$ these actions were lifted to conformally-flat backgrounds in \cite{KP19}.

Superconformal higher-spin (SCHS) gauge multiplets contain CHS fields at the component level. Hence, an effective method of studying various CHS models is to study the corresponding SCHS models which induce them. In the case of minimal depth CHS fields, such studies have already been initiated \cite{KMT, KP19, KPR}. However, the supersymmetric multiplets containing generalised CHS gauge fields have not yet appeared in the literature, neither at the superspace nor component level. It is therefore of interest to elaborate on generalised SCHS multiplets and their gauge invariant actions on curved backgrounds, which is the main subject of this paper. 

Another reason for the interest drawn by generalised CHS fields is their role in the context of the AdS/CFT correspondence. More specifically, it is known \cite{BG} that generalised CHS fields in $\mathbb{M}^d$ may be identified with the boundary values of partially massless fields \cite{DeserN1, DeserN2, Higuchi1, Higuchi2, Higuchi3, Metsaev2, DeserW1, DeserW2, DeserW3, DeserW4, Zinoviev, DNW  ,Vasiliev2006, Metsaev4,Brust} propagating in the bulk of AdS$_{d+1}$. In this work we will not comment further on these issues, however it should be mentioned that several publications on the supersymmetric partially massless side of this story have appeared recently \cite{Hinterbichler1,Hinterbichler2,Buchbinder, BHKP}.

This paper is organised as follows. In section \ref{section2} we present the generalised depth-$t$ superconformal gauge multiplets and derive their corresponding gauge and super-Weyl invariant actions on conformally-flat superspace backgrounds. Section \ref{section3} is devoted to extending gauge invariance of the action for the simplest supermultiplet containing the maximal depth conformal graviton to Bach-flat backgrounds. A new family of superconformal non-gauge multiplets is introduced in section \ref{section4} and their super-Weyl invariant actions are given. In addition to offering some concluding remarks, in section \ref{section5} we demonstrate that the model for the conformal hook field in a Bach-flat background is not determined uniquely by gauge and Weyl invariance. Throughout the entire paper we make use of conformal (super)space; appendix \ref{appendixA} summarises the relevant formulae. Appendix \ref{appendixB} discusses potential superspace realisations of the conformal hook models. 

%%%%%%%%%%%%%%%%%%%%%%%%%%%%%%%%%%%

\section{Generalised superconformal models} \label{section2}

To begin with, we recall the structure of $\mathcal{N}=1$ superconformal higher-spin multiplets 
\cite{KMT,KP19,KR}. For $n\geq m >0$, such a multiplet is 
formulated in terms of a prepotential $\Psi_{\a (n) \ad (m)} $, and its conjugate for $n\neq m$, defined modulo the gauge transformation
\bea
 \d_{ \xi, \eta} \Psi_{\a (n) \ad (m)} 
 =  
 \Nabla_{(\a_1}\xi_{\a_2 \dots \a_n)\ad(m)} 
+\bar \Nabla_{(\ad_1} \eta_{\a(n) \ad_2 \dots \ad_{m} )}~,
\label{1.3}
\eea
with unconstrained gauge parameters $\xi_{\a(m-1)\ad(n)}$ and $ \eta_{\a (m) \ad (n-1)} $. Here $\Nabla_A=(\Nabla_{a},\Nabla_{\a},\bNabla^{\ad})$ 
are the  covariant derivatives of $\mathcal{N}=1$ conformal superspace, see appendix \ref{appendixA} for more details. In the $n=m$ case one can consistently define the prepotential to be real, $H_{\a(n) \ad(n)}:=\J_{\a(n) \ad(n)} = \bar H_{\a(n) \ad(n)}$.\footnote{In Minkowski superspace, such gauge superfields with $n>1$ were introduced for the first time in \cite{HST}. The $n=1$ case corresponds to conformal supergravity \cite{FZ2}.}
For $n>m=0$, the prepotential  $\J_{\a(n)}$  is defined modulo the  gauge transformation
\bea
\label{ConfHSgravitino}
\d_{\xi,\l} \J_{\a(n)} = \Nabla_{(\a_{1}} \xi_{\a_2 \dots \a_n)} + \lambda_{\a{(n)}} ~, \qquad \bar{\Nabla}_{\ad} \l_{\a{(n)}} = 0 ~,
\eea
where the parameter $\xi_{\a(n-1)} $ is complex unconstrained, while $\l_{\a(n)}$ is covariantly 
chiral. Superconformal gauge-invariant actions for the multiplets \eqref{1.3} were constructed in \cite{KMT} 
in Minkowski superspace, while for arbitrary conformally flat backgrounds they were derived in \cite{KP19}.
Superconformal gauge-invariant actions for the multiplets \eqref{ConfHSgravitino} with $n>1$ were derived in \cite{KPR}
for conformally flat backgrounds. The action for the superconformal gravitino multiplet, which corresponds to $n=1$ in \eqref{ConfHSgravitino}, was described earlier  in Bach-flat backgrounds \cite{KMT} (see also \cite{KPR}). 

A specific feature of $\Psi_{\a (n) \ad (m)} $ with $n\geq m > 0$ is the presence of only a single spinor derivative in its gauge transformations \eqref{1.3}.
In this section we generalise these multiplets by increasing the number of spinor derivatives appearing in their gauge transformations.
%%%%%%%%%%%%%%%%%%%%%%%%%%%%%%%%%%%

\subsection{Generalised superconformal prepotentials and field strengths}

A depth-$t$ SCHS
multiplet $\U^{(t)}_{\a(n)\ad(m)}$, with $n\geq m \geq 1$, is defined modulo  gauge transformations
of the form
\begin{align}
\delta_{\zeta}\U^{(t)}_{\a(n)\ad(m)}=\big[\Nabla_{(\a_1},\bar{\Nabla}_{(\ad_1}\big]\Nabla_{\a_2\ad_2}\cdots\Nabla_{\a_t\ad_t}\zeta^{(t)}_{\a_{t+1}\dots\a_n)\ad_{t+1}\dots\ad_{m})}~,
 \label{depth}
\end{align}
where $\zeta^{(t)}_{\a(n-t)\ad(m-t)}$ is unconstrained and $1\leq t \leq \text{min}(n,m)$. 

If we require that both the prepotential and its corresponding gauge parameter are primary,
\begin{align}
K_B\U^{(t)}_{\a(n)\ad(m)}=0~,\qquad K_B\zeta^{(t)}_{\a(n-t)\ad(m-t)}=0~,
\end{align}
 then the Weyl weight and $\sU(1)_{R}$ charge carried by $\U^{(t)}_{\a(n)\ad(m)}$ must take the values
\begin{align}
\mathbb{D}\U^{(t)}_{\a(n)\ad(m)}=\big[t-\frac{1}{2}(n+m)\big]\U^{(t)}_{\a(n)\ad(m)}~,\qquad Y\U^{(t)}_{\a(n)\ad(m)}=\frac{1}{3}(n-m)\U^{(t)}_{\a(n)\ad(m)}~. \label{2.3}
\end{align}
For $n=m$ this allows us to choose both $\U^{(t)}_{\a(n)\ad(m)}$ and $\zeta^{(t)}_{\a(n-t)\ad(m-t)}$ to be real, in which case we will make use of the definition 
\begin{align}
H^{(t)}_{\a(n) \ad(n)} := \U^{(t)}_{\a(n) \ad(n)}=\bar{H}^{(t)}_{\a(n)\ad(n)}~. \label{2.4}
\end{align}

From the prepotential and its conjugate one may construct higher-derivative descendants 
\begin{subequations}
\begin{align}
\hat{\mathfrak{W}}_{\a(n+m-t+1)\ad(t-1)}^{(t)}(\U)&=\bigg(\bar{\Nabla}^{\bd_1}\Nabla_{(\a_1}-\frac{\text{i}t}{m-t+1}\Nabla_{(\a_1}{}^{\bd_1}\bigg)\Nabla_{\a_2}{}^{\bd_2}\cdots\Nabla_{\a_{m-t+1}}{}^{\bd_{m-t+1}} \label{Weyl1}\notag\\ 
&~~~\times\U_{\a_{m-t+2}\dots\a_{n+m-t+1})\ad(t-1)\bd(m-t+1)}~, \\
\check{\mathfrak{W}}_{\a(n+m-t+1)\ad(t-1)}^{(t)}(\bar{\U})&=\bigg(\bar{\Nabla}^{\bd_1}\Nabla_{(\a_1}-\frac{\text{i}t}{n-t+1}\Nabla_{(\a_1}{}^{\bd_1}\bigg)\Nabla_{\a_2}{}^{\bd_2}\cdots\Nabla_{\a_{n-t+1}}{}^{\bd_{n-t+1}} \label{Weyl2}\notag\\
&~~~\times\bar{\U}_{\a_{n-t+2}\dots\a_{n+m-t+1})\ad(t-1)\bd(n-t+1)}~.
\end{align}
\end{subequations}
They are the higher-depth analogues of the linearised higher-spin super-Weyl tensors since they both prove be primary in a generic background
\begin{align}
K_B\hat{\mathfrak{W}}_{\a(n+m-t+1)\ad(t-1)}^{(t)}(\U)=0~, \qquad K_B \check{\mathfrak{W}}_{\a(n+m-t+1)\ad(t-1)}^{(t)}(\bar{\U})=0~.
\end{align}
In addition, when restricted to conformally-flat backgrounds,
\begin{align}
W_{\a(3)}=0 \label{ConfFBBG}~
\end{align}
where $W_{\a(3)}$ is the super-Weyl tensor (see appendix \ref{appendixA} for more details), it is possible to show that they are invariant under the higher-depth gauge transformations \eqref{depth},
\begin{align}
\delta_{\zeta}\hat{\mathfrak{W}}_{\a(n+m-t+1)\ad(t-1)}^{(t)}(\U)=0~, \qquad \delta_{\zeta} \check{\mathfrak{W}}_{\a(n+m-t+1)\ad(t-1)}^{(t)}(\bar{\U})=0~.
\end{align}

In principle one can instead consider supermultiplets $\hat{\U}^{(t)}_{\a(n)\ad(m)}$ with the gauge freedom
\begin{align}
\delta_{\hat{\zeta}}\hat{\U}^{(t)}_{\a(n)\ad(m)}=\Nabla_{(\a_1}\bNabla_{(\ad_1}\Nabla_{\a_2\ad_2}\cdots\Nabla_{\a_t\ad_t}\hat{\zeta}^{(t)}_{\a_{t+1}\dots\a_{n})\ad_{t+1}\dots\ad_m)}~.
\end{align}
In order for $\hat{\U}^{(t)}_{\a(n)\ad(m)}$ to be primary it must have Weyl weight $\big(t-\frac{1}{2}(n+m)\big)$ and $\sU(1)_{R}$ charge $\frac{1}{3}\big(n-m+2t\big)$. In this case one may consistently define $\hat{\U}^{(t)}_{\a(n)\ad(m)}$ to be longitudinal anti-linear, $\Nabla_{(\a_1}\hat{\U}^{(t)}_{\a_2\dots\a_{n+1)}\ad(m)}=0$. However, these supermultiplets possess several undesirable features. In particular, it is not possible to impose a reality condition similar to \eqref{2.4}, nor does it seem possible to construct gauge and super-Weyl invariant actions describing their dynamics. It is for this reason that we will focus only on the supermultiplets $\U^{(t)}_{\a(n)\ad(m)}$.

%%%%%%%%%%%%%%%%%%%%%%%%%%%%%%%%%%%

\subsection{Generalised superconformal actions}

Using the field strengths \eqref{Weyl1} and \eqref{Weyl2}, one may construct the action functional
\begin{align}
S_{\text{Skeleton}}^{(n,m,t)}[\U,\bar{\U}]=\text{i}^{n+m+2}\int\text{d}^{4|4}z\, E \, \hat{\mathfrak{W}}^{\a(n+m-t+1)\ad(t-1)}_{(t)}(\U)\check{\mathfrak{W}}_{\a(n+m-t+1)\ad(t-1)}^{(t)}(\bar{\U}) +\text{c.c.}~, \label{HDaction}
\end{align}
which is locally superconformal (on any background) and gauge invariant on conformally-flat backgrounds. The overall coefficient of $\text{i}^{n+m+2}$ in \eqref{HDaction} has been chosen because of the identity
\begin{align}
\text{i}^{n+m+1}\int\text{d}^{4|4}z\, E \, \hat{\mathfrak{W}}^{\a(n+m-t+1)\ad(t-1)}_{(t)}(\U)\check{\mathfrak{W}}_{\a(n+m-t+1)\ad(t-1)}^{(t)}(\bar{\U}) +\text{c.c.} ~=~ 0~,
\end{align}
 which holds in conformally-flat backgrounds up to an irrelevant total derivative.
 
The action \eqref{HDaction} may be recast into the alternative forms
\begin{align}
S_{\text{Skeleton}}^{(n,m,t)}[\U,\bar{\U}]&=\text{i}^{n+m+2}\int\text{d}^{4|4}z\, E \, \bar{\U}^{\a(m)\ad(n)}_{(t)}\hat{\mathfrak{B}}_{\a(m)\ad(n)}^{(t)}(\U) +\text{c.c.} \notag\\
&= \text{i}^{n+m+2}\int\text{d}^{4|4}z\, E \,\U^{\a(n)\ad(m)}_{(t)}\check{\mathfrak{B}}_{\a(n)\ad(m)}^{(t)}(\bar{\U}) +\text{c.c.}
\end{align}
where we have made use of the following definitions 
\begin{subequations}
\begin{align}
\hat{\mathfrak{B}}_{\a(m)\ad(n)}^{(t)}(\U)&= \bigg(\Nabla^{\b_1}\bar{\Nabla}_{(\ad_1}-\frac{\text{i}t}{n-t+1}\Nabla_{(\ad_1}{}^{\b_1}\bigg)\Nabla_{\ad_2}{}^{\b_2}\cdots\Nabla_{\ad_{n-t+1}}{}^{\b_{n-t+1}}
\notag\\
&\phantom{=}\times\hat{\mathfrak{W}}^{(t)}_{\b(n-t+1)\a(m)\ad_{n-t+2}\dots\ad_n)}(\U)~, \label{Bach1}\\
\check{\mathfrak{B}}_{\a(n)\ad(m)}^{(t)}(\bar{\U})&=\bigg(\Nabla^{\b_1}\bar{\Nabla}_{(\ad_1}-\frac{\text{i}t}{m-t+1}\Nabla_{(\ad_1}{}^{\b_1}\bigg)\Nabla_{\ad_2}{}^{\b_2}\cdots\Nabla_{\ad_{m-t+1}}{}^{\b_{m-t+1}}
\notag\\
&\phantom{=}\times\check{\mathfrak{W}}^{(t)}_{\b(m-t+1)\a(n)\ad_{m-t+2}\dots\ad_m)}(\bar{\U})~. \label{Bach2}
\end{align}
\end{subequations}
 The two tensors \eqref{Bach1} and \eqref{Bach2} are primary in a generic background. Additionally, in any conformally-flat background, they may be shown to possess the following properties:
\begin{enumerate}
\item  Both $\hat{\mathfrak{B}}_{\a(m)\ad(n)}^{(t)}$ and $\check{\mathfrak{B}}_{\a(n)\ad(m)}^{(t)}$ are invariant under the gauge transformations \eqref{depth},
\begin{subequations}
\begin{align}
\delta_{\zeta}\hat{\mathfrak{B}}_{\a(m)\ad(n)}^{(t)}(\U)=0~,\qquad\delta_{\zeta}\check{\mathfrak{B}}_{\a(n)\ad(m)}^{(t)}(\bar{\U})=0~.
\end{align}
\item Both $\hat{\mathfrak{B}}_{\a(m)\ad(n)}^{(t)}$ and $\check{\mathfrak{B}}_{\a(n)\ad(m)}^{(t)}$ are `partially conserved',
\begin{align}
0&=\big[\Nabla^{\b_1},\bar{\Nabla}^{\bd_1}\big]\Nabla^{\b_2\bd_2}\cdots\Nabla^{\b_t\bd_t}\hat{\mathfrak{B}}_{\b(t)\a(m-t)\bd(t)\ad(n-t)}^{(t)}(\U)~,\\
0&=\big[\Nabla^{\b_1},\bar{\Nabla}^{\bd_1}\big]\Nabla^{\b_2\bd_2}\cdots\Nabla^{\b_t\bd_t}\check{\mathfrak{B}}_{\b(t)\a(n-t)\bd(t)\ad(m-t)}^{(t)}(\bar{\U})~.
\end{align} 
\item $\hat{\mathfrak{B}}_{\a(m)\ad(n)}^{(t)}$ and $\check{\mathfrak{B}}_{\a(n)\ad(m)}^{(t)}$ are related by complex conjugation as follows,\footnote{This relation contains non-trivial information. Its direct check is quite time consuming. }
\begin{align} 
\overline{\hat{\mathfrak{B}}}^{(t)}_{\a(n)\ad(m)}(\bar{\U}):=\overline{\big(\hat{\mathfrak{B}}_{\a(m)\ad(n)}^{(t)}(\U)\big)}&=\check{\mathfrak{B}}^{(t)}_{\a(n)\ad(m)}(\bar{\U})~.
\end{align}
\end{subequations}
\end{enumerate}
On account of these properties, both \eqref{Bach1} and \eqref{Bach2} may be interpreted as the higher-depth analogues of the higher-spin linearised super-Bach tensors given in \cite{KPR}.

%%%%%%%%%%%%%%%%%%%%%%%%%%%%%%%%%%%%%

\subsection{Wess-Zumino gauge for minimal depth supermultiplets}
\label{WZsubsec}

For the supermultiplets with depth $t>1$, the presence of vector derivatives in the gauge transformations \eqref{depth} prevents one from constructing a Wess-Zumino gauge. Consequently, all component fields are present in this case, some of which are depth $t-1$, $t$ and $t+1$ CHS gauge fields. However in the special case $t=1$, when the depth assumes its minimal value,\footnote{Unlike the non-supersymmetric case, minimal depth SCHS fields do not correspond to the ordinary SCHS fields described at the beginning of this section (see eq. \eqref{1.3}). } this is no longer an obstruction and the multiplet shortens. It is therefore of interest to elaborate on the construction of the corresponding Wess-Zumino gauge. 

We restrict our analysis to bosonic backgrounds \eqref{Bbackground} which are conformally-flat \eqref{ConfFBBG}. Together these conditions imply that the (bosonic) Weyl tensor $C_{abcd}$ vanishes 
\bea
\quad W_{\a(3)}  = 0 \quad\implies\quad C_{abcd}=0 ~.
\eea
The minimal depth supermultiplets, which we hereby denote $\U_{\a(n) \ad(m)}:=\U^{(1)}_{\a(n) \ad(m)}$, have Weyl weight and $\sU(1)_R$ charge given by
\bea
\mathbb{D} \U_{\a(n) \ad(m)} = \big[ 1 - \frac{1}{2} (n+m)  \big] \U_{\a(n) \ad(m)} ~, \quad Y \U_{\a(n) \ad(m)} = \frac{1}{3} (n-m) \U_{\a(n) \ad(m)} ~.
\eea
Additionally, they are defined modulo the gauge freedom
\begin{align}
\delta_{\zeta}\U_{\a(n)\ad(m)}=\big[\Nabla_{(\a_1},\bar{\Nabla}_{(\ad_1}\big]\zeta_{\a_{2}\dots\a_n)\ad_{2}\dots\ad_{m})}~. \label{depth1}
\end{align}

By making use of \eqref{depth1}, one may choose a Wess-Zumino gauge such that there are twelve non-vanishing primary component fields: 
\begin{subequations} \label{2.16}
\bea
\psi_{\a(n+1)\ad(m)} &=& \Nabla_{(\a_1} \U_{\a_2 \dots \a_{n+1}) \ad(m)}| ~, \label{2.16a} \\
\o_{\a(n)\ad(m+1)} &=& \bNabla_{(\ad_1} \U_{\a(n) \ad_2 \dots \ad_{m+1})}| ~, \label{2.16b} \\
A_{\a(n) \ad(m)} &=& - \frac{1}{4} \Nabla^2 \U_{\a(n) \ad(m)} | ~, \label{2.16c} \\
B_{\a(n) \ad(m)} &=& - \frac{1}{4} \bNabla^2 \U_{\a(n) \ad(m)} | ~, \label{2.16d} \\
h_{\a(n+1) \ad(m+1)} &=& \frac{1}{2} \big[ \Nabla_{(\a_1} , \bNabla_{(\ad_1} \big] \U_{\a_2 \dots \a_{n+1}) \ad_2 \dots \ad_{m+1})} | ~, \label{2.16e} \\
\chi_{\a(n+1) \ad(m-1)} &=& \frac{1}{2} \big[\Nabla_{(\a_1} , \bNabla^{\bd} \big] \U_{\a_2 \dots \a_{n+1}) \ad(m-1) \bd} | ~, \label{2.16f}\\
\r_{\a(n-1) \ad(m+1)} &=& \frac{1}{2} \big[\Nabla^\b , \bNabla_{(\ad_1} \big] \U_{\a(n-1) \b \ad_2 \dots \ad_{m+1} )} | ~, \label{2.16g}\\
\varphi_{\a(n+1)\ad(m)} &=& - \frac{1}{4} \Nabla_{(\a_1} \bNabla^2 \U_{\a_2 \dots \a_{n+1}) \ad(m)} | + \frac{\ri}{2}\frac{2m+1 }{m} \nabla_{(\a_1}{}^{\bd} \o_{\a_2 \dots \a_{n+1}) \ad(m) \bd} ~,\label{2.16h}\\
\g_{\a(n) \ad(m+1)} &=& -\frac{1}{4} \bNabla_{(\ad_1} \Nabla^2 \U_{\a(n) \ad_2 \dots \ad_{m+1})}| - \frac{\ri}{2}\frac{2n+1 }{n} \nabla^{\b}{}_{(\ad_1} \psi_{\a(n) \b \ad_2 \dots \ad_{m+1})} ~, \label{2.16i}\\
\Phi_{\a(n) \ad(m-1)} &=& - \frac{1}{4} \bNabla^{\bd} \Nabla^2 \U_{\a(n) \ad(m-1) \bd}| - \frac{\ri}{2}\frac{2m+1 }{m+n+1} \nabla^{\bb} \psi_{\a(n) \b \ad(m-1) \bd} ~, \label{2.16j}\\
\G_{\a(n-1) \ad(m)} &=& - \frac{1}{4} \Nabla^{\b} \bNabla^2 \U_{\a(n-1) \b \ad(m)} | + \frac{\ri}{2}\frac{2n+1 }{m+n+1} \nabla^{\bb} \o_{\a(n-1) \b \ad(m) \bd}~, \label{2.16k}\\
V_{\a(n) \ad(m)} &=& \frac{1}{32} \{ \Nabla^2 , \bNabla^2 \} \U_{\a(n) \ad(m)} | + \frac{\ri}{4}\frac{n}{m} \nabla^{\b}{}_{(\ad_1} \chi_{\a(n) \b \ad_2 \dots \ad_m)} \non \\
&& - \frac{\ri}{4}\frac{m}{n} \nabla_{(\a_1}{}^{\bd} \r_{\a_2 \dots \a_n) \ad(m) \bd} ~.\label{2.16l}
\eea
\end{subequations}

For all values of $n$ and $m$ the gauge transformation laws for the first five fields of \eqref{2.16} are
\begin{subequations}
\label{gtWZ}
\bea
\d_{\e} \psi_{\a(n+1) \ad(m)} &=& \nabla_{(\a_1 (\ad_1} \e_{\a_2 \dots \a_{n+1}) \ad_2 \dots \ad_m)} ~, \\
\d_{\eta} \o_{\a(n) \ad(m+1)} &=& \nabla_{(\a_1 (\ad_1} \eta_{\a_2 \dots \a_{n}) \ad_2 \dots \ad_{m+1})} ~, \\
\d_{\l} A_{\a(n) \ad(m)} &=& \nabla_{(\a_1 (\ad_1} \l_{\a_2 \dots \a_n) \ad_2 \dots \ad_m)} ~, \\
\d_{\m} B_{\a(n) \ad(m)} &=& \nabla_{(\a_1 (\ad_1} \m_{\a_2 \dots \a_n) \ad_2 \dots \ad_m)} ~, \\
\d_{\xi} h_{\a(n+1) \ad(m+1)} &=& \nabla_{(\a_1 (\ad_1} \nabla_{\a_2 \ad_2} \xi_{\a_3 \dots \a_{n+1}) \ad_3 \dots \ad_{m+1})} ~,
\eea
\end{subequations}
revealing that \eqref{2.16a}--\eqref{2.16d}
are all depth-1 CHS fields, whilst \eqref{2.16e}
is a depth-2 CHS field. 

In general, the class (i.e. generalised CHS, conformal non-gauge\footnote{See section \ref{section4} for further explanation of non-gauge conformal fields.} etc.) and residual gauge transformations of the remaining component fields \eqref{2.16f} - \eqref{2.16l} depends upon the values of $n$ and $m$. We will now consider each of the distinct cases in turn.
 It should be noted that for $n=m$, the reality condition \eqref{2.4} may be imposed and, as a result, the component structure \eqref{2.16} simplifies significantly. In particular,
\begin{subequations}
\begin{align}
\bar{\o}_{\a(n+1) \ad(n)} &= \psi_{\a(n+1) \ad(n)}~,\qquad \qquad \bar{\g}_{\a(n+1) \ad(n)} = \varphi_{\a(n+1) \a(n)} ~, \\
\bar{B}_{\a(n) \ad(n)} &= A_{\a(n) \ad(n)}~, \qquad\qquad ~~\bar{\G}_{\a(n) \ad(n-1)} = \Phi_{\a(n) \ad(n-1)} ~,  \\
\bar{h}_{\a(n+1) \ad(n+1)} &= h_{\a(n+1) \ad(n+1)} ~, \qquad \qquad \bar{V}_{\a(n) \ad(n)} = V_{\a(n) \ad(n)} ~,\\
\bar{\r}_{\a(n+1) \ad(n-1)} &= \chi_{\a(n+1) \ad(n-1)} ~. 
\end{align}
\end{subequations}
The special case $n=m=1$  will be examined separately, and in more detail, in section \ref{WZMDSpin2}.

For $n > m=1$, we find that $\r_{\a(n-1) \ad(2)}$ and $\g_{\a(n) \ad(2)}$ are depth-1 and depth-2 CHS fields
\begin{subequations}
\bea
\d_{\t} \r_{\a(n-1) \ad(2)} &=& \nabla_{(\a_1 (\ad_1} \t_{\a_2 \dots \a_{n-1}) \ad_2)} ~,\\
\d_{\theta} \g_{\a(n) \ad(2)} &=& \nabla_{(\a_1 (\ad_1} \nabla_{\a_2 \ad_2)} \theta_{\a_3 \dots \a_n) } ~,
\eea
\end{subequations}
whilst $\chi_{\a(n+1)}$, $\varphi_{\a(n+1) \ad}$, $\Phi_{\a(n)}$, $\G_{\a(n-1) \ad}$ and $V_{\a(n) \ad}$ are non-gauge fields.

Finally, if $n\geq m > 1$  then $\r_{\a(n-1)\ad(m+1)}$ and $\chi_{\a(n+1)\ad(m-1)}$ have depth-1, whilst $\gamma_{\a(n)\ad(m+1)}$, $\vf_{\a(n+1)\ad(m)}$ and $V_{\a(n)\ad(m)}$ have depth-2,
\begin{subequations}
\bea
\d_{\t} \r_{\a(n-1) \ad(m+1)} &=& \nabla_{(\a_1 (\ad_1} \t_{\a_2 \dots \a_{n-1}) \ad_2\dots\ad_{m+1})} ~,\\
\d_{\theta} \g_{\a(n) \ad(m+1)} &=& \nabla_{(\a_1 (\ad_1} \nabla_{\a_2 \ad_2} \theta_{\a_3 \dots \a_n) \ad_3\dots\ad_{m+1})} ~,\\
\d_{\s} \chi_{\a(n+1) \ad(m-1)} &=& \nabla_{(\a_1 (\ad_1} \s_{\a_2 \dots\a_{n+1}) \ad_2 \dots \ad_{m-1})} ~, \\
\d_{\k} \varphi_{\a(n+1) \ad(m)} &=& \nabla_{(\a_1 (\ad_1} \nabla_{\a_2 \ad_2} \k_{\a_3\dots\a_{n+1})\ad_3 \dots \ad_{m})} ~,\\
\d_{\ell} V_{\a(n) \ad(m)} &=& \nabla_{(\a_1 (\ad_1} \nabla_{\a_2 \ad_2} \ell_{\a_3 \dots \a_n) \ad_3 \dots \ad_{m})} ~.
\eea
\end{subequations} 
The only non-gauge fields in this case are $\Phi_{\a(n) \ad(m-1)}$ and $\G_{\a(n-1) \ad(m)}$.

%%%%%%%%%%%%%%%%%%%%%%%%%%%%%%%%%%%%%%%%%%%%%%%%%%%%%%%
%%%%%%%%%%%%%%%%%%%%%%%%%%%%%%%%%%%%%%%%%%%%%%%%%%%%%%%

\section{Maximal-depth conformal graviton supermultiplet} \label{section3}

The maximal-depth conformal graviton supermultiplet is described by the weightless real primary superfield\footnote{In this section we drop all labels referring to the depth since we deal only with $t=1$.} $H_{\a\ad}$ which is inert under $\sU(1)_{R}$ transformations,
\begin{subequations}\label{gengrav}
\begin{align}
K_{B}H_{\a\ad}=0~,\qquad \mathbb{D}H_{\a\ad}=0~,\qquad YH_{\a\ad}=0~,\qquad  H_{\a\ad}=\bar{H}_{\a\ad}~,
\end{align}
 and which is defined modulo the depth-1 gauge transformations
\begin{align}
\delta_{\zeta}H_{\a\ad}=\big[\nabla_{\a},\bar{\nabla}_{\ad}\big]\zeta~. \label{gt2}
\end{align}
\end{subequations}
Here the real gauge parameter $\zeta$ is primary and unconstrained. 
This corresponds to the generalised supermultiplet of section \ref{section2} with $n=m=1$ and is called the maximal-depth conformal graviton supermultiplet because, as will be shown shortly, it contains the maximal-depth conformal graviton at the component level.

%%%%%%%%%%%%%%%%%%%%%%%%%%%%%

\subsection{Gauge invariant action in Bach-flat background}

The linearised super-Weyl tensor associated with $H_{\a\ad}$, and its corresponding gauge variation under \eqref{gt2}, is given by
\begin{subequations}
\begin{align}
\mathfrak{W}_{\a(2)}(H)&=\bigg(\bar{\Nabla}^{\bd}\Nabla_{(\a_1}-\text{i}\Nabla_{(\a_1}{}^{\bd}\bigg)H_{\a_2)\bd} ~,\\
\delta_{\zeta}\mathfrak{W}_{\a(2)}(H)&= 3\bigg(2W_{\a(2)}{}^{\b}\Nabla_{\b}\zeta+\Nabla_{\b}W_{\a(2)}{}^{\b}\zeta\bigg)~.
\end{align}
\end{subequations}
 It follows that the skeleton action
\begin{align}
\label{HSkeleton}
S_{\text{Skeleton}}[H]=\frac{1}{3}\int \text{d}^{4|4} z\, E \, \mathfrak{W}^{\a(2)}(H)\mathfrak{W}_{\a(2)}(H) +\text{c.c.}
\end{align}
(here we have chosen a different overall normalisation as compared to \eqref{HDaction}) has gauge variation proportional to the background super-Weyl tensor 
\begin{align}
\delta_{\zeta}S_{\text{Skeleton}}[H]=2\int\text{d}^{4|4}z\, E \, \zeta\bigg\{3\Nabla_{\g}W^{\g\a(2)}\mathfrak{W}_{\a(2)}(H)-2W^{\g\a(2)}\Nabla_{\g}\mathfrak{W}_{\a(2)}(H)\bigg\} +\text{c.c.}
\end{align}

It is possible to restore gauge invariance to the skeleton by supplementing it with the non-minimal primary action
\begin{align}
S_{\text{NM}}[H]=\int\text{d}^{4|4}z\, E \, H^{\a\ad}W_{\a}{}^{\b(2)}\Nabla_{\b}H_{\b\ad}+\text{c.c.}
\end{align}  
The action which is gauge invariant in a Bach-flat background may then be shown to be
\begin{align}
S[H]=S_{\text{Skeleton}}[H]-3S_{\text{NM}}[H]~,\qquad \delta_{\zeta}S[H]\bigg|_{B_{\a\ad}=0}=0~, \label{3.77}
\end{align}
where $B_{\a\ad}$ is the super-Bach tensor \eqref{super-Bach}. 

%%%%%%%%%%%%%%%%%%%%%%%%%%%%%%%%%

\subsection{The component action}
\label{WZMDSpin2}

In this subsection we employ the Wess-Zumino gauge constructed in section \ref{WZsubsec} for the purpose of reducing \eqref{3.77} to components. While this gauge fixing was realised only on conformally-flat backgrounds \eqref{ConfFBBG}, what follows applies more generally to backgrounds satisfying \eqref{Bbackground}.

Examining \eqref{2.16} we find that in the Wess-Zumino gauge $H_\aa$ contains seven non-vanishing primary fields
\begin{subequations}
\label{HComponents}
\bea
\psi_{\a(2)\ad} &=& \Nabla_{(\a_1} H_{\a_2) \ad}| ~, \\
A_{\aa} &=& - \frac{1}{4} \Nabla^2 H_{\aa} | ~, \\
h_{\a(2) \ad(2)} &=& \frac{1}{2} \big[ \Nabla_{(\a_1} , \bNabla_{(\ad_1} \big] H_{\a_2) \ad_2)} | ~, \\
\chi_{\a(2)} &=& \frac{1}{2} \big[\Nabla_{(\a_1} , \bNabla^{\ad} \big] H_{\a_2) \ad} | ~, \\
\varphi_{\a(2)\ad} &=& - \frac{1}{4} \Nabla_{(\a_1} \bNabla^2 H_{\a_2) \ad} | + \frac{3 \ri}{2} \nabla_{(\a_1}{}^{\bd} \bar{\psi}_{\a_2) \ad \bd} ~, \\
\Phi_\a &=& - \frac{1}{4} \bNabla^{\ad} \Nabla^2 H_{\aa}| - \frac{\ri}{2} \nabla^{\bb} \psi_{\a \b \bd} ~, \\
V_{\aa} &=& \frac{1}{32} \{ \Nabla^2 , \bNabla^2 \} H_{\aa} | + \frac{\ri}{4} \nabla^{\b}{}_{\ad} \chi_{\a \b} - \frac{\ri}{4} \nabla_{\a}{}^{\bd} \bar{\chi}_{\ad \bd} ~.
\eea
\end{subequations}
Both $h_{\a(2)\ad(2)}$ and $V_{\a\ad}$ are real whilst all other component fields are complex. Associated with \eqref{HComponents} are the following gauge fixing conditions
\begin{subequations}
\label{Hgfc}
\bea
\big[ \Nabla_{\a} , \bNabla_{\ad} \big] \z | &=& 0 ~, \\
\bNabla_{\ad} \Nabla^2 \z | &=& 3 \ri \nabla_{\aa} \Nabla^{\a} \z | =: \frac{3}{2} \nabla_{\aa} \e^{\a} ~, \\
\{ \Nabla^2 , \bNabla^2 \} \z | &=& - 4 \Box \z | =: 2 \Box \xi ~.
\eea
\end{subequations}
The residual gauge transformations \eqref{gtWZ} are generated by the fields $\l := - \frac{1}{4} \Nabla^2 \z|$, $\e_{\a}$ and $\xi$
\begin{subequations}
	\bea
	\d_{\l} A_\aa &=& \nabla_{\aa} \l ~, \\
	\d_{\e} \psi_{\a(2) \ad} &=& \nabla_{(\a_1 \ad} \e_{\a_2)} ~, \\
	\d_\xi h_{\a(2) \ad(2)} &=& \nabla_{(\a_1 (\ad_1} \nabla_{\a_2) \ad_2)} \xi ~. \label{333}
	\eea
\end{subequations}

Using the above definitions, the action \eqref{3.77} may be readily reduced to components
\bea
\label{MaxDepthSpin2ComponentAction}
S[H] &=& - \int \rd^4 x \, e\, \bigg\{ \frac{3}{4} \bigg[\mathfrak{C}^{\a(3) \ad}(h) \mathfrak{C}_{\a(3) \ad}(h) -h^{\a(2)\ad(2)}C_{\a(2)}{}^{\b(2)}h_{\b(2)\ad(2)}\bigg] \non\\
&+& \frac{3 \ri}{2} \bigg[\hat{\mathfrak{C}}^{\a(3)} (\psi) \check{\mathfrak{C}}_{\a(3)} (\bar{\psi})-\psi^{\a(2)\ad}\bigg(C_{\a(2)}{}^{\b(2)}\nabla_{\b}{}^{\bd}\bar{\psi}_{\b\ad\bd}-\nabla_{\b}{}^{\bd}C_{\a(2)}{}^{\b(2)}\bar{\psi}_{\b\ad\bd}\bigg)\bigg]  \non\\
&+& 2 \hat{\mathfrak{C}}^{\a(2)} (A) \check{\mathfrak{C}}_{\a(2)} (\bar{A}) - 2 \ri \bar{\varphi}^{\a \ad(2)} \mathfrak{X}_{\a \ad(2)}(\varphi) + \frac{1}{4} \bar{\chi}^{\ad(2)} \mathfrak{X}_{\ad(2)} (\chi) - \frac{ \ri}{2} \bar{\Phi}^{\ad} \mathfrak{X}_\ad(\Phi)\non \\
&-& 2 V^{\aa} V_{\aa} \bigg\} + \text{c.c.} \\
&\equiv& - \frac{3}{4} S[h] + \frac{3}{2} S[\psi, \bar{\psi}] + 8 S[A,\bar{A}] - 2  S[\varphi , \bar{\varphi}] + \frac{1}{4} S[\chi,\bar{\chi}] + \frac{1}{2} S[\Phi, \bar{\Phi}] - 2 S[V] \non ~. 
\eea
This action has been expressed in a manifestly conformal and gauge invariant (in a Bach-flat background) form by using the field strengths associated with each conformal field,
\begin{subequations}
\begin{align}
\hat{\mathfrak{C}}_{\a(3)} (\psi)&=\nabla_{(\a_1}{}^{\bd}\psi_{\a_2\a_3)\bd}~, \qquad\qquad\quad \mathfrak{C}_{\a(3)\ad}(h)=\nabla_{(\a_1}{}^{\bd}h_{\a_2\a_3)\ad\bd} ~,\\
\check{\mathfrak{C}}_{\a(3)} (\bar{\psi})&=\nabla_{(\a_1}{}^{\bd_1}\nabla_{\a_2}{}^{\bd_2}\bar{\psi}_{\a_3)\bd(2)}~,\qquad ~\mathfrak{X}_{\ad(2)}(\chi)=\nabla_{(\ad_1}{}^{\b_1}\nabla_{\ad_2)}{}^{\b_2}\chi_{\b(2)} ~,\\
\hat{\mathfrak{C}}_{\a(2)} (A)&=\nabla_{(\a_1}{}^{\bd}A_{\a_2)\bd}~, \qquad\qquad\quad \phantom{..}\mathfrak{X}_{\a \ad(2)}(\varphi)=\nabla_{(\ad_1}{}^{\b}\vf_{\b\a\ad_2)}~,\\
\check{\mathfrak{C}}_{\a(2)} (\bar{A})&=\nabla_{(\a_1}{}^{\bd}\bar{A}_{\a_2)\bd}~, \qquad \qquad \qquad ~\phantom{..} \mathfrak{X}_\ad(\Phi)=\nabla_{\ad}{}^{\b}\Phi_{\b}~, 
\end{align}
\end{subequations}
along with the non-minimal counter terms necessary for gauge invariance.

The analysis above indicates that the component action decomposes into a (diagonal) sum of gauge invariant actions -- denoted $S[h], S[\psi,\bar{\psi}]$ and $S[A,\bar{A}]$ -- describing a maximal-depth conformal graviton $h_{\a(2)\ad(2)}$ \cite{KP19}, a conformal gravitino $\psi_{\a(2)\ad}$ and a complex Maxwell field $A_{\a\ad}$ respectively.  In addition, there are also several non-gauge fields \cite{KPR} $\chi_{\a(2)}, \vf_{\a(2)\ad}$ and $\Phi_{\a}$ present, the latter of which describes a massless Weyl spinor. The vector field $V_\aa$ is auxiliary and is only present to ensure off-shell supersymmetry.

Upon degauging and converting to vector notation, the action $S[h]$ for the depth-2 conformal graviton in \eqref{MaxDepthSpin2ComponentAction} may be shown to to be proportional to (see \cite{KP19-2} for more details)
\begin{align}
 S[h]\propto\int \text{d}^4x \, e \,\bigg\{&h^{ab}\Box h_{ab}-\frac{4}{3}\mathcal{D}_{a}h^{ab}\mathcal{D}^{c}h_{bc}-2R_{ab}h^{ac}h_{c}{}^{b}+\frac{1}{6}Rh^{ab}h_{ab}+2C_{abcd}h^{ac}h^{bd}\bigg\}~. \label{788.9}
 \end{align}
 Here $\mathcal{D}_a$ is the torsion-free Lorentz covariant derivative and 
 $h_{ab}$ is symmetric and traceless. This action is invariant under the (degauged version of the) gauge transformations \eqref{333}
\begin{align}
\delta_{\xi}h_{ab}=\big(\mathcal{D}_a\mathcal{D}_b-\frac{1}{2}R_{ab}\big)\xi-\frac{1}{4}\eta_{ab}\big(\Box-\frac 12 R\big)\xi~. \label{88.88}
\end{align} 
This action has appeared in various forms 
over the past forty years, see in particular \cite{DeserN1, DeserN2,  EO, Sachs, BT2015, DeserW6, KP19}. We refer the reader to \cite{KP19-2} for a more thorough account of its history.  

For a more detailed analysis on the gauge invariant action $S[\psi,\bar{\psi}]$  for the conformal gravitino on Bach-flat backgrounds see \cite{KP19}. The overall sign of the action \eqref{HDaction} has been chosen so that the Maxwell action in \eqref{MaxDepthSpin2ComponentAction},
\begin{align}
S[A,\bar{A}]=- \frac{1}{2} \int \text{d}^4x \, e \, F^{ab}(A)F_{ab}(\bar{A})~,\qquad F_{ab}(A):=\mathcal{D}_{a}A_{b}-\mathcal{D}_{b}A_{a}~,
\end{align}
comes with canonical sign. 

Finally, we would like to point out that the action \eqref{3.77} may be recast into the form
\begin{align}
S[H]=\frac{3}{2}\int \text{d}^{4|4}z H^{\a\ad}&\bigg\{\frac{3}{2}D^{\b}\bar{D}^2D_{\b}H_{\a\ad}-\frac{1}{2}\big[D_{\a},\bar{D}_{\ad}\big]\big[D_{\b},\bar{D}_{\bd}\big]H^{\b\bd} \non\\
&~-4\partial_{\a\ad}\partial^{\b\bd}H_{\b\bd}-\frac{1}{4}\big\{D^2,\bar{D}^2\big\} H_{\a\ad}\bigg\}
\label{3.15}
\end{align}
 in Minkowski superspace and defines a superconformal field theory.
The action is invariant under the gauge transformation 
$\delta_{\zeta}H_{\a\ad}=\big[D_{\a},\bar{D}_{\ad}\big]\zeta$, which is the flat-superspace form of  \eqref{gt2}.
There exists a model for linearised supergravity constructed in \cite{BGLP} with a larger gauge freedom
\bea
\delta H_{\a\ad}=\big[D_{\a},\bar{D}_{\ad}\big]\zeta + \l_{\a\ad} + \bar \l_{\a\ad}~, 
\qquad \bar D_\bd \l_{\a\ad} =0~,
\eea
than that which the action \eqref{3.15} possesses. However, the $\l$ gauge symmetry proves to be incompatible with 
the superconformal invariance. 

%%%%%%%%%%%%%%%%%%%%%%%%%%%%%%%%%%%%%

\section{Superconformal non-gauge models} \label{section4}

In Bach-flat backgrounds, conformal non-gauge fields\footnote{Conformal non-gauge fields were first described in $\mathbb{M}^d$ by Vasiliev \cite{Vasiliev} and on curved backgrounds in \cite{ KPR}.} $\chi_{\a(n)\ad(m)}$  play an essential role in ensuring gauge invariance in models for the following three CHS fields: (i) conformal maximal-depth spin-3 \cite{KP19-2}; (ii) conformal maximal-depth spin-5/2\cite{KP19-2}; and (iii) conformal (minimal-depth) hook field\cite{KPR}. 

Common to all three of these models is the presence of non-gauge fields $\chi_{\a(n)}$ with $m=0$. Such fields may be found sitting within the so-called chiral non-gauge supermultiplets $\Omega_{\a(n)}$ (reviewed below). The latter were first introduced in \cite{KPR}, where $\Omega_{\a}$ played an important role in ensuring gauge invariance of the supersymmetric extension of (iii). However, non-gauge fields with $m> 0$ were also important in models for (i) and (ii), and these are not contained within $\Omega_{\a(n)}$ at the component level. This motivates the search for superconformal non-gauge multiplets containing $\chi_{\a(n)\ad(m)}$ for any $n$ and $m$.

\subsection{Chiral supermultiplets}

Chiral non-gauge superfields and their corresponding kinetic actions were proposed in \cite{KPR}. For convenience we now recall the main elements of these models. 

 A primary non-gauge chiral superfield $\Omega_{\a(n)}$, with $n\geq 1$, satisfies 
\begin{subequations}\label{7.11+12}
\begin{align}
K_B\Omega_{\a(n)}=0~,\qquad \bar{\Nabla}_{\ad}\Omega_{\a(n)}=0~, \label{7.11}
\end{align}
 Consistency of these two conditions with the superconformal algebra demands that the Weyl weight and $\sU(1)_{R}$ charge of $\Omega_{\a(n)}$ are related by 
\begin{align}
\mathbb{D}\Omega_{\a(n)}=\Delta\Omega_{\a(n)}~, \qquad
Y\Omega_{\a(n)}=-\frac{2}{3}\Delta\Omega_{\a(n)}~. \label{7.12}
\end{align}
\end{subequations}
If we choose $\Delta=1-\frac{1}{2}n$, then it can be shown that the composite scalar superfield defined by
\begin{align}
\mathcal{F}^{(n)}\big(\Omega,\bar{\Omega}\big)=&\sum_{k=0}^{n}(-1)^k\Nabla_{\a_1\ad_1}\cdots\Nabla_{\a_k\ad_k}\Omega^{\a(n)}\Nabla_{\a_{k+1}\ad_{k+1}}\cdots\Nabla_{\a_n\ad_n}\bar{\Omega}^{\ad(n)} \notag\\
-\frac{\text{i}}{2}&\sum_{k=1}^{n}(-1)^{n+k}\Nabla_{\a_1}\Nabla_{\a_2\ad_2}\cdots\Nabla_{\a_k\ad_k}\Omega^{\a(n)}\bar{\Nabla}_{\ad_1}\Nabla_{\a_{k+1}\ad_{k+1}}\cdots\Nabla_{\a_n\ad_n}\bar{\Omega}^{\ad(n)}
\label{F-Lagrangian}
\end{align}
is primary in a generic background. The superconformal properties of $\mathcal{F}^{(n)}$ may therefore be summarised as follows
\begin{align}
K_{A}\mathcal{F}^{(n)}=0~,\qquad \mathbb{D}\mathcal{F}^{(n)}=2\mathcal{F}^{(n)}~,\qquad Y\mathcal{F}^{(n)}=0~.
\end{align} 
Furthermore, one can show that it satisfies the complex conjugation property 
\begin{align}
\overline{\mathcal{F}^{(n)}}=(-1)^n\mathcal{F}^{(n)}~. 
\end{align}
It follows that the action functional
\begin{align}
S_{\text{Chiral}}^{(n)}[\Omega,\bar{\Omega}]=\text{i}^{n}\int \text{d}^{4|4}z\, E \, \mathcal{F}^{(n)}\big(\Omega,\bar{\Omega}\big) \label{7.16}
\end{align}
is real and super-Weyl invariant. 
When written as an integral over the chiral subspace, this action simplifies to
\begin{align}
S_{\text{Chiral}}^{(n)}[\Omega,\bar{\Omega}]=-\frac{~\text{i}^{n}}{4}\int \rd^4x \rd^2 \q \, \cE\, \O^{\a(n)}\bar{\Nabla}^2\Nabla_{\a_1\ad_1}\cdots\Nabla_{\a_n\ad_n}\bar{\O}^{\ad(n)}~.
\end{align}

\subsection{Longitudinal linear supermultiplets} \label{section4.2}

A superfield $\Omega_{\a(n)\ad(m)}$, with $n \geq m$, is said to be longitudinal linear if it obeys the constraint
\begin{align}
 \bar{\Nabla}_{(\ad_1}\Omega_{\a(n)\ad_2\dots\ad_{m+1})}=0~\quad \implies \quad \bNabla^2\Omega_{\a(n)\ad(m)}=0~.
\end{align}

Similar to the chiral case, requiring $\Omega_{\a(n)\ad(m)}$ to be primary fixes its $\sU(1)_{R}$ charge in terms of its conformal weight as follows
\begin{align}
\mathbb{D}\Omega_{\a(n)\ad(m)}=\Delta \Omega_{\a(n)\ad(m)}~, \qquad Y\Omega_{\a(n)\ad(m)}=-\frac{2}{3}(\Delta+m)\Omega_{\a(n)\ad(m)}~.
\end{align}
Choosing $\Delta=1-\frac{1}{2}(n-m)$ allows one to construct the following superconformal action
\begin{align}
S_{~||}^{(n,m)}[\Omega,\bar{\Omega}]=\text{i}^{m+n}\int\text{d}^{4|4}z\, E \, \mathcal{F}^{(n,m)}(\Omega,\bar{\Omega})~,\label{long}
\end{align}
where $\mathcal{F}^{(n,m)}(\Omega,\bar{\Omega})$ is the composite scalar superfield
\begin{align}
\mathcal{F}^{(n,m)}(\Omega,\bar{\Omega})=&\sum_{k=0}^{n-m}(-1)^k\Nabla_{\ad_1}{}^{\b_1}\cdots\Nabla_{\ad_k}{}^{\b_k}\bar{\Omega}^{\a(m)\ad(n)}\notag\\[-12pt]
&\phantom{\sum_{k=0}^{n-m}(-1)^k}\times\Nabla_{\ad_{k+1}}{}^{\b_{k+1}}\cdots\Nabla_{\ad_{n-m}}{}^{\b_{n-m}}\Omega_{\a(m)\b(n-m)\ad_{n-m+1}\dots\ad_{n}} \notag\\[-10pt]
-\frac{\text{i}}{2}&\sum_{k=1}^{n-m}(-1)^{n+m+k}\bar{\Nabla}_{\ad_1}\Nabla_{\ad_2}{}^{\b_2}\cdots\Nabla_{\ad_k}{}^{\b_k}\bar{\Omega}^{\a(m)\ad(n)}\notag\\[-12pt]
&\phantom{\sum_{k=0}^{n-m}(-1)^k}\times\Nabla^{\b_1}\Nabla_{\ad_{k+1}}{}^{\b_{k+1}}\cdots\Nabla_{\ad_{n-m}}{}^{\b_{n-m}}\Omega_{\a(m)\b(n-m)\ad_{n-m+1}\dots\ad_{n}}~,
\end{align}
possessing the properties
\begin{subequations}
\begin{align}
K_{A}\mathcal{F}^{(n,m)}=0~,&\qquad \mathbb{D}\mathcal{F}^{(n,m)}=2\mathcal{F}^{(n,m)}~,\qquad Y\mathcal{F}^{(n,m)}=0~, \\[5pt]
&~~~ \overline{\mathcal{F}^{(n,m)}}=(-1)^{n+m}\mathcal{F}^{(n,m)}~. 
\end{align}
\end{subequations}
 When written in the chiral subspace the action \eqref{long} takes the form
\begin{align}
S_{~||}^{(n,m)}&[\Omega,\bar{\Omega}]=-\frac{\text{i}^{m+n}}{4}\int\text{d}^{4}x\text{d}^2\theta \,  \cE \, \bigg\{ \Omega_{\a(m)\b(n-m)\bd(m)}\bar{\Nabla}^2\Nabla_{\ad_1}{}^{\b_1}\cdots\Nabla_{\ad_{n-m}}{}^{\b_{n-m}}\bar{\Omega}^{\a(m)\ad(n-m)\bd(m)} \notag\\
&+\frac{2m}{m+1}\Nabla_{\ad_1}{}^{\b_1}\cdots\Nabla_{\ad_{n-m}}{}^{\b_{n-m}}\bar{\Nabla}_{\dd}\bar{\Omega}^{\a(m)\ad(n-m)\dd\bd(m-1)}\bar{\Nabla}^{\gd}\Omega_{\a(m)\b(n-m)\bd(m-1)\gd} \bigg\}~. 
\end{align}

The non-vanishing independent component fields of $\Omega_{\a(n)\ad(m)}$ are defined according to
\begin{subequations}\label{4.13}
\begin{align}
A_{\a(n)\ad(m)}&:=\Omega_{\a(n)\ad(m)}| ~,\label{4.13a}\\
B_{\a(n-1)\ad(m)}&:= \Nabla^{\b}\Omega_{\a(n-1)\b\ad(m)}|~, \label{4.13b}\\
C_{\a(n+1)\ad(m)}&:= \Nabla_{(\a_1}\Omega_{\a_2\dots\a_{n+1})\ad(m)}|~,\label{4.13c}\\
D_{\a(n)\ad(m-1)}&:= \bNabla^{\bd}\Omega_{\a(n)\ad(m-1)\bd}|~,\label{4.13d}\\
E_{\a(n)\ad(m)}&:=-\frac{1}{4}\Nabla^2\Omega_{\a(n)\ad(m)}|~,\label{4.13e}\\
F_{\a(n-1)\ad(m-1)}&:=\frac{1}{2}\big[\Nabla^{\b},\bNabla^{\bd}\big]\Omega_{\a(n-1)\b\ad(m-1)\bd}|+\text{i}\frac{m+1}{n+1}\Nabla^{\b\bd}\Omega_{\a(n-1)\b\ad(m-1)\bd}|~,\label{4.13f}\\
G_{\a(n+1)\ad(m-1)}&:=\frac{1}{2}\big[\Nabla_{(\a_1},\bNabla^{\bd}\big]\Omega_{\a_2\dots\a_{n+1})\ad(m-1)\bd}|~,\label{4.13g}\\
H_{\a(n)\ad(m-1)}&:= -\frac{1}{4}\bNabla^{\bd}\Nabla^2\Omega_{\a(n)\ad(m-1)\bd}|+\text{i}\frac{m-n}{m}\Nabla^{\b\bd}\Nabla_{(\b}\Omega_{\a_1\dots\a_{n})\ad(m-1)\bd}|~.\label{4.13h}
\end{align}
\end{subequations}
 
 The first two fields, \eqref{4.13a} and \eqref{4.13b}, are primary and have the same conformal weight as that of a maximal depth CHS field of the same rank (though the former do not have any gauge symmetry). The next four fields \eqref{4.13c} -- \eqref{4.13f}
 are all primary and are conformal non-gauge. However, the last two fields \eqref{4.13g} and \eqref{4.13h} are not able to be defined so that they are primary. Instead they transform non-trivially under a $K$-transformation, 
\begin{subequations}
\begin{align}
K_{\b\bd}G_{\a(n+1)\ad(m-1)}&=8\text{i}(m+1)\ve_{\b(\a_1}A_{\a_2\dots\a_{n+1})\ad(m-1)\bd}~, \\
K_{\b\bd}H_{\a(n)\ad(m-1)}&=-4\text{i}n\frac{m+1}{n+1}\ve_{\b(\a_1}B_{\a_2\dots\a_n)\ad(m-1)\bd}~,
\end{align} 
\end{subequations} 
and do not correspond to typical (i.e. generalised CHS or non-gauge) conformal fields.\footnote{See, however, \cite{Metsaev, Metsaev3} where various conformal fields were defined to transform non-trivially under special conformal transformations.}

Here we do not give the corresponding component action, since it is not illuminating. Rather it suffices to give a few comments regarding its structure. First, by setting $m=0$ in the above models, one recovers the rank-$n$ chiral non-gauge models from the previous section. Being chiral, the component content of these supermultiplets is simple and there are only four non-vanishing fields, all of which turn out to be conformal and non-gauge \cite{KPR}. The component action is also simple in the sense that it consists only of the kinetic terms for the four non-gauge fields and is diagonal. However, the longitudinal linear supermultiplets have twice as many component fields, and not all of them are primary but instead transform into one another under Weyl transformations. Thus, in order to maintain Weyl invariance, the component action necessarily becomes non-diagonal, resulting in a much more complicated structure.

\section{Discussion} \label{section5}

It has been conjectured that lower-spin conformal fields are neccesary in ensuring gauge invariance of minimal depth CHS fields on Bach-flat backgrounds \cite{GrigorievT}.  Indeed, in support of this proposal, there have appeared various fully worked examples of CHS models (with fields of varying depth) where a coupling between the parent CHS field and subsidiary conformal non-gauge fields were crucial for gauge invariance\cite{KP19-2,KPR}.  Therefore, we expect that the longitudinal linear non-gauge supermultiplets, presented in section \ref{section4.2}, will play an equally important role in ensuring the gauge invariance of various SCHS fields on super-Bach flat backgrounds.\footnote{In such models it would be interesting to better understand the role of the non-primary component fields \eqref{4.13g} and \eqref{4.13h} present in the longitudinal linear supermultiplets.}   

Of the few existing examples of complete gauge invariant models, those describing conformal gauge fields with depth greater than one constitute the majority. This is because their construction is more tractable as compared to their minimal depth cousins, on account of their lower-derivative skeletons. In this paper we have described, for the first time, the supersymmetric analogues of these generalised (i.e. higher-depth) CHS gauge fields and their gauge invariant actions on conformally-flat backgrounds. In doing so we have initiated a program to investigate their Bach-flat completions, beginning with the maximal depth graviton supermultiplet detailed in section \ref{section3}. 
This supermultiplet is the lowest rank member of a family of depth one supermultiplets. The latter are particularly interesting because they have shortened multiplets and contain a collection of depth one  (i.e. ordinary) and depth two CHS fields, as well as non-gauge conformal fields. 

Understanding the ingredients that are necessary in constructing gauge invariant models for (S)CHS fields (of all depths) in Bach-flat backgrounds is an important technical problem. Moreover, given the necessity of subsidiary fields, there is no reason to expect that the requirements of gauge and Weyl invariance should determine the Bach-flat completion of a generic (S)CHS model uniquely. Indeed, we now give an example of such a scenario, and demonstrate that there exists an infinite family of Bach-flat completions for the conformal pseudo-graviton (also known as a traceless hook field).

The latter is described by the field $h_{\a(3)\ad}$, and its conjugate $\bar{h}_{\a\ad(3)}$, possessing the properties
\begin{align}
K_{\b\bd}h_{\a(3)\ad}&=0~,\qquad \mathbb{D}h_{\a(3)\ad}=0~, \notag\\
\delta_{\ell}&h_{\a(3)\ad}=\nabla_{(\a_1\ad}\ell_{\a_2\a_3)}~.
\end{align}
The sector consisting of only the pseudo-graviton is given by
\begin{align}
S[h,\bar{h};\Gamma]=S_{\text{Skeleton}}[h,\bar{h}]+\Gamma S_{\text{NM},1}[h,\bar{h}]-\frac{1}{2}S_{\text{NM},2}[h,\bar{h}]\label{hook0}
\end{align}
where $\Gamma\in\mathbb{R}$ is a free parameter and\footnote{The linearised Weyl tensors corresponding to the pseudo-graviton take the form $\hat{\mathfrak{C}}_{\a(4)}(h)=\nabla_{(\a_1}{}^{\bd}h_{\a_2\a_3\a_4)\bd}$ and $\check{\mathfrak{C}}_{\a(4)}(\bar{h})=\nabla_{(\a_1}{}^{\bd_1}\nabla_{\a_2}{}^{\bd_2}\nabla_{\a_3}{}^{\bd_3}\bar{h}_{\a_4)\bd(3)}$.} 
\begin{subequations}
\begin{align}
S_{\text{Skeleton}}[h,\bar{h}]&= \int\text{d}^4x\, e \, \hat{\mathfrak{C}}^{\a(4)}(h)\check{\mathfrak{C}}_{\a(4)}(\bar{h})+\text{c.c.}~, \\
S_{\text{NM},1}[h,\bar{h}]&= \int\text{d}^4x\, e \, h^{\a(3)\ad}C_{\a(3)}{}^{\b}\bar{C}_{\ad}{}^{\bd(3)}\bar{h}_{\b\bd(3)} +\text{c.c.}~,\\
S_{\text{NM},2}[h,\bar{h}]&= \int\text{d}^4x\, e \,h^{\a(3)\ad}\bigg\{5C_{\a(3)}{}^{\g}\nabla_{\g}{}^{\bd}\nabla^{\b\bd}\bar{h}_{\b\bd(2)\ad}+6C_{\a(2)}{}^{\g(2)}\nabla_{\g}{}^{\bd}\nabla_{\g}{}^{\bd}\bar{h}_{\a\ad\bd(2)}\notag\\
& -6\nabla_{\g}{}^{\bd}C_{\a(2)}{}^{\b\g}\nabla_{\a}{}^{\bd}\bar{h}_{\b\bd(2)\ad}+2\nabla^{\d\bd}C_{\a(3)}{}^{\b}\nabla_{\d}{}^{\bd}\bar{h}_{\b\bd(2)\ad}-4\nabla^{\b\bd}\nabla_{\g}{}^{\bd}C_{\a(3)}{}^{\g}\bar{h}_{\b\bd(2)\ad} \notag\\
&+\nabla_{\g}{}^{\bd}C_{\a(3)}{}^{\g}\nabla^{\b\bd}\bar{h}_{\b\bd(2)\ad}\bigg\} +\text{c.c.}
\end{align}
\end{subequations}

The action \eqref{hook0} is gauge invariant up to terms quadratic in the Weyl tensor, which is why  $\Gamma$ remains free at this stage. In order to ensure gauge invariance to all orders, it is necessary to introduce some other conformal field transforming non-trivially under the pseudo-graviton gauge transformations. In \cite{KPR} use was made of the non-gauge field $\chi_{\a(2)}$ with the properties
\begin{subequations}
\begin{align}
K_{\b\bd}\chi_{\a(2)}=0&~, \qquad \mathbb{D}\chi_{\a(2)}=\chi_{\a(2)}~, \\
\delta_{\ell}\chi_{\a(2)}&=C_{\a(2)}{}^{\b(2)}\ell_{\b(2)}~.
\end{align}
\end{subequations} 
Then the action which is gauge invariant in a Bach-flat background takes the form 
\begin{align}
S_{\text{Hook}}[h,\chi]=S[h,\bar{h};\Gamma=1]-2S[h,\chi]+S[\chi,\bar{\chi}]~, \label{hook1}
\end{align}
where 
\begin{subequations}
\begin{align}
S[h,\bar{\chi}]&=\int\text{d}^4x\, e \, h^{\a(3)\ad}\bigg\{C_{\a(3)}{}^{\g}\nabla_{\g}{}^{\bd}\bar{\chi}_{\bd\ad}-\nabla_{\g}{}^{\bd}C_{\a(3)}{}^{\g}\bar{\chi}_{\bd\ad}\bigg\}+\text{c.c.}~, \\
S[\chi,\bar{\chi}]&= \int\text{d}^4x\, e \, \bar{\chi}^{\ad(2)}\nabla_{\ad}{}^{\a}\nabla_{\ad}{}^{\a}\chi_{\a(2)}+\text{c.c.}
\end{align}
\end{subequations}

However, it turns out that this action is not unique, and one can instead use a different non-gauge field $\varphi_{\a(4)\ad(2)}$, with the properties
\begin{subequations}
\begin{align}
K_{\b\bd}\varphi_{\a(4)\ad(2)}=0~,& \qquad \mathbb{D}\varphi_{\a(4)\ad(2)}=\varphi_{\a(4)\ad(2)}~, \\
\delta_{\ell}\varphi_{\a(4)\ad(2)}&=C_{\a(4)}\bar{\ell}_{\ad(2)}~, \label{hihello}
\end{align}
\end{subequations} 
to achieve gauge invariance. In this case, the gauge invariant action takes the form
\begin{align}
S_{\text{Hook}}[h,\vf]=S[h,\bar{h};\Gamma=3]+2S[h,\vf]-S[\vf,\bar{\vf}]~, \label{hook2}
\end{align}
where 
\begin{subequations}
\begin{align}
S[h,\vf]&= \int\text{d}^4x\, e \, h^{\a(3)\ad}\bigg\{\bar{C}_{\ad}{}^{\gd\bd(2)}\nabla_{\gd}{}^{\b}\vf_{\a(3)\b\bd(2)}-\nabla_{\gd}{}^{\b}\bar{C}_{\ad}{}^{\gd\bd(2)}\vf_{\a(3)\b\bd(2)}\bigg\}+\text{c.c.}~, \\
S[\vf,\bar{\vf}]&=\int\text{d}^4x\, e \, \bar{\vf}^{\a(2)\ad(4)}\nabla_{\ad}{}^{\a}\nabla_{\ad}{}^{\a}\vf_{\a(4)\ad(2)}+\text{c.c.}
\end{align}
\end{subequations}

In fact, by using both of the fields $\chi_{\a(2)}$ and $\vf_{\a(4)\ad(2)}$, one can construct a one-parameter family of gauge invariant actions for the pseudo-graviton described by  
\begin{align}
S_{\text{Hook}}[h,\chi,\vf;\Gamma]=S[h,\bar{h};\Gamma]~+~&(\Gamma-3)S[h,\bar{\chi}]-\frac{1}{2}(\Gamma-3)S[\chi,\bar{\chi}] \notag\\
~+~&(\Gamma-1)S[h,\vf]-\frac{1}{2}(\Gamma-1)S[\vf,\bar{\vf}]~.~~~~~~~~~ \label{hook3}
\end{align}

The actions \eqref{hook1} and \eqref{hook2} may be recovered by setting $\Gamma=1$ and $\Gamma=3$ respectively. In appendix \ref{appendixB} we illustrate the utility of the longitudinal linear non-gauge supermultiplets  by proposing various ways that one can realise supersymmetric extensions of the new one-parameter family of models \eqref{hook3}.  

Finally, it would be interesting to investigate the possible non-linear completions of the model \eqref{788.9} for the generalised conformal graviton using cohomological techniques along the lines of \cite{Boulanger1, Boulanger2}. Given the similarity of the generalised conformal graviton to the partially massless graviton, and the various no-go theorems regarding self-interactions of the latter (see for instance \cite{deRham, Joung, Rosen} and references therein), the supermultiplet \eqref{gengrav} could play an important role in such an analysis (cf. the discussions in \cite{Hinterbichler1, Boulanger3}).  

%%%%%%%%%%%%%%%%%%%%%%%%%%%%%%%%%%%%%%%%%%%%%%%%%%%
%%%%%%%%%%%%%%%%%%%%%%%%%%%%%%%%%%%%%%%%%%%%%%%%%%%

\noindent
{\bf Acknowledgements:}\\
The work of SMK is supported in part by the Australian 
Research Council, project No. DP200101944.
The work of MP and ESNR is supported by the Hackett Postgraduate Scholarship UWA,
under the Australian Government Research Training Program.

\appendix

\section{$\mathcal{N}=1$ conformal superspace in four dimensions}\label{appendixA}

In this appendix we collate the elements of the conformal superspace approach to conformal supergravity which are essential to this work.
We refer the reader to the original paper \cite{ButterN=1} for more details (see also appendix A of \cite{KPR}). In this work we adopt the spinor conventions of \cite{BK}.

We consider a curved $\cN=1$ superspace $\mathcal{M}^{4|4}$
parametrised by local coordinates 
$z^{M} = 
(x^{m},\theta^{\m},\bar \theta_{\dot{\mu}})$.  
The structure group is chosen to be $\sSU(2,2|1)$ and thus the covariant derivatives 
$\Nabla_A$
have the form
\begin{align}
\Nabla_A &= (\Nabla_a, \Nabla_\alpha, \bar\Nabla^\ad)
=E_A{}^M \pa_M - \hf \Omega_A{}^{bc} M_{bc} - \ri \Phi_A Y
- B_A \mathbb{D} - \mathfrak{F}_{A}{}^B K_B ~,
\label{6.1}
\end{align}
where
$\Omega_A{}^{bc}$  
denotes the Lorentz connection,  $\Phi_A$  the  $\rm U(1)_R$ connection, $B_A$
the dilatation connection, and  $\mathfrak F_A{}^B$ the special
superconformal connection.

Below we list the graded commutation relations for the $\cN=1$ superconformal 
algebra following the conventions
adopted in \cite{KPR}. We note that the translation generators 
$P_A = (P_a, Q_\a ,\bar Q^\ad)$ have been replaced with $\Nabla_A$ and all (anti-)commutators not listed vanish.

The Lorentz generators act on vectors and Weyl spinors as follows:
	\bea
	M_{ab} V_{c} = 2 \eta_{c[a} V_{b]} ~, \qquad 
	M_{\a \b} \j_{\g} = \ve_{\g (\a} \j_{\b)} ~, \qquad \bar{M}_{\ad \bd} \bar \j_{\gd} = \ve_{\gd ( \ad} \bar \j_{\bd )} ~.
	\eea
The  $\rm U(1)_R$, dilatation and special conformal generators obey
\begin{subequations}
	\begin{align}
	[Y, \Nabla_\a] &= \Nabla_\a ~,\quad [Y, \bar\Nabla^\ad] = - \bar\Nabla^\ad~,   \\
	[\mathbb{D}, \Nabla_a] &= \Nabla_a ~, \quad
	[\mathbb{D}, \Nabla_\a] = \hf \Nabla_\a ~, \quad
	[\mathbb{D}, \bar\Nabla^\ad ] = \hf \bar\Nabla^\ad ~\\
	[Y, S^\a] &= - S^\a ~, \quad
	[Y, \bar{S}_\ad] = \bar{S}_\ad~, \quad \{ S_\a , \bar{S}_\ad \} = 2 \ri  K_{\aa} \\
	[\mathbb{D}, K_a] &= - K_a ~, \quad
	[\mathbb{D}, S^\a] = - \hf S^\a~, \quad
	[\mathbb{D}, \bar{S}_\ad ] = - \hf \bar{S}_\ad ~.
	\end{align}
 The algebra of $K^A$ and $\Nabla_B$ takes the form
\begin{align}
[K_\aa, \Nabla_\bb] &= 4 \big(\ve_{\ad \bd} M_{\a \b} +  \ve_{\a \b} \bar{M}_{\ad \bd} -  \ve_{\a \b} \ve_{\ad \bd} \mathbb{D} \big) ~, \\
\{ S_\a , \Nabla_\b \} &= \ve_{\a \b} \big( 2 \mathbb{D} - 3 Y \big) - 4 M_{\a \b} ~, \\
\{ \bar{S}_\ad , \bar{\Nabla}_\bd \} &= - \ve_{\ad \bd} \big( 2 \mathbb{D} + 3 Y) + 4 \bar{M}_{\ad \bd}  ~, \\
[K_{\a \ad}, \Nabla_\b] &= - 2 \ri \ve_{\a \b} \bar{S}_{\ad} \ , \qquad \qquad \qquad[K_\aa, \bar{\Nabla}_\bd] =
2 \ri  \ve_{\ad \bd} S_{\a} ~,  \\
[S_\a , \Nabla_\bb] &= 2 \ri \ve_{\a \b} \bar{\Nabla}_{\bd} \ , \qquad \qquad \quad \qquad[\bar{S}_\ad , \Nabla_\bb] =
- 2 \ri \ve_{\ad \bd} \Nabla_{\b} ~.
\end{align}
\end{subequations}

In conformal superspace, 
the torsion and  curvature tensors are subject to covariant constraints such that 
$[\Nabla_A, \Nabla_B\}$ are expressible solely in terms of the super-Weyl tensor
$W_{\alpha \beta \gamma}= W_{(\a\b\g)}$. The latter is a primary chiral 
superfield of dimension 3/2 and $\rm U(1)_R$ charge $-1$, 
\bea
K_B W_{\a \b \g} =0~, \quad \bar \Nabla_\bd W_{\a\b\g}=0 ~, \quad 
{\mathbb D} W_{\a\b\g} = \frac 32 W_{\a\b\g}~, \quad YW_{\a\b\g}=-W_{\a\b\g}~. \label{Weylboy}
\eea
The solutions to the aforementioned constraints are given by
\begin{subequations}
	\label{CSSAlgebra}
	\bea
	\{ \Nabla_{\a} , \Nabla_{\b} \} & = & 0 ~, \quad \{ \bar{\Nabla}_{\ad} , \bar{\Nabla}_{\bd} \} = 0 ~, \quad \{\Nabla_{\a} , \bar{\Nabla}_{\ad} \} = - 2 \ri \Nabla_{\a \ad} ~, \\
	\big[ \Nabla_{\a} , \Nabla_{\b \bd} \big] & = & \ri \ve_{\a \b} \Big( 2 \bar{W}_{\bd}{}^{\gd \dd} \bar{M}_{\gd \dd} - \frac{1}{2} \bar{\Nabla}^{\ad} \bar{W}_{\ad \bd \gd} \bar{S}^{\gd} + \frac{1}{2} \Nabla^{\g \ad} \bar{W}_{\ad \bd}{}^{\gd} K_{\g \gd} \Big) ~, \\
	\big[ \bar{\Nabla}_{\ad} , \Nabla_{\b \bd} \big] & = & - \ri \ve_{\ad \bd} \Big( 2 W_{\b}{}^{\g \d} M_{\g \d} + \frac{1}{2} \Nabla^{\a} W_{\a \b \g} S^{\g} + \frac{1}{2} \Nabla^{\a \gd} W_{\a \b}{}^{\g} K_{\g \gd} \Big) ~,\\
	\big[ \Nabla_{\a \ad} , \Nabla_{\b \bd} \big] & = & \ve_{\ad \bd} \bigg(W_{\a \b}{}^{\g} \Nabla_{\g} + \Nabla^{\g} W_{\a \b}{}^{\d} M_{\g \d} - \frac{1}{8} \Nabla^{2} W_{\a \b \g} S^{\g} + \frac{\ri}{2} \Nabla^{\g \gd} W_{\a \b \g} \bar{S}_{\gd} \non \\
	&& + \frac{1}{4} \Nabla^{\g \dd} \Nabla_{(\a} W_{\b) \g}{}^{\d} K_{\d \dd} + \frac{1}{4} \Nabla^{\g} W_{\a \b \g}\big( 2\mathbb{D} - 3Y\big)\bigg)\notag\\
	& -&\ve_{\a \b} \bigg(\bar{W}_{\ad \bd}{}^{\gd} \bar{\Nabla}_{\gd} + \bar{\Nabla}^{\gd} \bar{W}_{\ad \bd}{}^{\dd} \bar{M}_{\gd \dd} - \frac{1}{8} \bar{\Nabla}^{2} \bar{W}_{\ad \bd \gd} \bar{S}^{\gd} - \frac{\ri}{2} \Nabla^{\g \gd} \bar{W}_{\ad \bd \gd} S_{\g} \non \\
	&& + \frac{1}{4} \Nabla^{\d \gd} \bar{\Nabla}_{(\ad} \bar{W}_{\bd) \gd}{}^{\dd} K_{\d \dd} + \frac{1}{4} \bar{\Nabla}^{\gd} \bar{W}_{\ad \bd \gd}\big(2 \mathbb{D} +3Y\big)\bigg)
	~. \label{2.8d}
	\eea
\end{subequations}
We also find that $W_{\a \b \g}$ obeys the Bianchi identity
\bea
B_{\a\ad} :=  \ri \Nabla^\b{}_{\ad} \Nabla^\g W_{\a\b\g}
=\ri \Nabla_{\a}{}^{ \bd} \bar \Nabla^\gd \bar W_{\ad\bd\gd}
= \bar B_{\a\ad}~,
\label{super-Bach}
\eea
where the primary superfield
$B_{\a\ad}$ is the super-Bach tensor and was introduced in \cite{BK88}.

If we restrict our attention to bosonic backgrounds, defined by
\bea
\Nabla_{a}| = \nabla_a~, \quad W_{\a(3)}| = 0~, \quad \Nabla_{\a} W_{\b(3)}| = - C_{\a \b(3)} ~, \quad \Nabla^2 W_{\a(3)}| = 0 ~, \label{Bbackground}
\eea
then the resulting geometry describes conformal gravity. Here, the only surviving component field of $B_{\a \ad}$ is the Bach tensor $B_{\a(2)\ad(2)}$
\bea
\label{Bach}
-\frac{1}{2} \big[ \Nabla_{(\a_1} , \bar{\Nabla}_{(\ad_1}\big] B_{\a_2) \ad_2)} | =  \nabla_{(\ad_1}{}^{\b_1}\nabla_{\ad_2)}{}^{\b_2}C_{\a(2)\b(2)}=:B_{\a(2) \ad(2)} ~,
\eea
where  $C_{\a\b\g\d}$  is the anti self-dual part of the Weyl tensor  $C_{abcd}$. Now, the geometry of spacetime is described by the conformally covariant derivative
\begin{align}
\nabla_{a}=e_{a}{}^{m}\partial_m-\frac{1}{2}\omega_{a}{}^{bc}M_{bc}-\mathfrak{b}_a\mathbb{D}-\mathfrak{f}_{a}{}^{b}K_b \label{10.1}
\end{align}
 which obeys the algebra
\bea \label{10.2}
\big[\nabla_{\a\ad},\nabla_{\b\bd} \big]&=&-\big(\ve_{\ad\bd}C_{\a\b\g\d}M^{\g\d}+\ve_{\a\b}\bar{C}_{\ad\bd\gd\dd}\bar{M}^{\gd\dd}\big) \notag\\
&&
-\frac{1}{4}\big(\ve_{\ad\bd}\nabla^{\d\gd}C_{\a\b\d}{}^{\g}+\ve_{\a\b}\nabla^{\g\dd}\bar{C}_{\ad\bd\dd}{}^{\gd}\big)K_{\g\gd}~.
\eea
 
 \section{Superspace realisations of the conformal hook model} \label{appendixB}

In section \ref{section5} we derived a one-parameter family of gauge invariant models for the conformal hook field. In this appendix we  
propose various superspace realisations of these models.   
 
  By analysing the resulting component structure, it was shown in \cite{KPR} that the model \eqref{hook3} with $\Gamma=1$ (i.e. \eqref{hook1}) could be embedded within a supersymmetric gauge invariant action for a superfield $\U_{\a(2)}$ coupled to a chiral non-gauge superfield $\Omega_{\a}$. In this model, the latter two transformed according to
\begin{subequations} 
\begin{align}
\delta_{\zeta,\lambda}\U_{\a(2)}=\Nabla_{(\a_1}&\zeta_{\a_2)}+\lambda_{\a(2)}~,\qquad \bar{\Nabla}_{\ad}\lambda_{\a(2)}=0~, \label{d}\\
&\delta_{\lambda} \Omega_{\a}=W_{\a}{}^{\b(2)}\lambda_{\b(2)}~,
\end{align} 
\end{subequations}
 with $\zeta_{\a}$ unconstrained and $\lambda_{\a(2)}$ covariantly chiral. 
 
For $\Gamma\neq 3$, it may be possible to embed \eqref{hook3} within a one-parameter family of supersymmetric actions by introducing, in addition to $\U_{\a(2)}$ and $\Omega_{\a}$, the longitudinal non-gauge field $\Omega_{\a(3)\ad(2)}$. The latter should be entangled with $\U_{\a(2)}$ through the transformations
 \begin{align}
 \delta_{\zeta}\Omega_{\a(3)\ad(2)}=W_{\a(3)}\bar{\Nabla}_{(\ad_1}\bar{\zeta}_{\ad_2)}~.\label{a} 
 \end{align}
 Then, the component field of $\Omega_{\a(3)\ad(2)}$ defined by \eqref{4.13c} corresponds to $\vf_{\a(4)\ad(2)}$ and has the correct gauge transformations \eqref{hihello} (in an appropriate Wess-Zumino gauge).
 
When $\Gamma=3$, the non-gauge field $\chi_{\a(2)}$ is no longer present in the action \eqref{hook3}. However, as explained in \cite{KPR}, $\chi_{\a(2)}$ appears as a component field of the gauge superfield $\U_{\a(2)}$. Hence, at the component level, any gauge invariant model for $\U_{\a(2)}$ will necessarily contain a coupling between $h_{\a(3)\ad}$ and $\chi_{\a(2)}$. Therefore  $\U_{\a(2)}$ is not a suitable supermultiplet in providing a supersymmetric embedding for the action \eqref{hook2}. 

A possible candidate for this role\footnote{One can rule out the superfield $\U_{\a(3)}$ (which also contains $h_{\a(3)\ad}$) as there is no possible coupling to a non-gauge superfield which contains $\vf_{\a(4)\ad(2)}$ at the component level and which preserves all symmetries.} is the spin-5/2 supermultiplet $\U_{\a(2)\ad}$ coupled to a longitudinal non-gauge field $\Omega_{\a(4)\ad(2)}$ according to\footnote{We would like to point out that the gauge transformations \eqref{a} and \eqref{b} preserve the gauge-for-gauge symmetries $\zeta_{\a}\rightarrow \zeta_{\a}+\Nabla_{\a}\sigma$ and $\Lambda_{\a\ad}\rightarrow \Lambda_{\a\ad}+\Nabla_{\a}\bar{\eta}_{\ad}$ of \eqref{d} and \eqref{c}, respectively.}
\begin{subequations}
\begin{align}
\delta_{\zeta,\Lambda}\U_{\a(2)\ad}&=\bar{\Nabla}_{\ad}\zeta_{\a(2)}+\Nabla_{(\a_1}\Lambda_{\a_2)\ad}\label{c} ~,\\
\delta_{\Lambda}\Omega_{\a(4)\ad(2)}&=W_{(\a_1\a_2\a_3}\bar{\Nabla}_{(\ad_1}\bar{\Lambda}_{\a_4)\ad_2)}~.\label{b}
\end{align}
\end{subequations}
with unconstrained and complex gauge parameters $\zeta_{\a(2)}$ and $\Lambda_{\a\ad}$. In this scenario, it is the component field \eqref{4.13e} of $\Omega_{\a(4)\ad(2)}$ which may be identified with the non-gauge field $\vf_{\a(4)\ad(2)}$.

%%%%%%%%%%%%%%%%%%%%%%%%%%%%%%%%%%%%%%%%%%%%%%%%%%%%%%%%%%%%%%%
%%%%%%%%%%%%%%%%%%%%%%%%%%%%%%%%%%%%%%%%%%%%%%%%%%%%%%%%%%%%%%%
%%%%%%%%%%%%%%%%%%%%%%%%%%%%%%%%%%%%%%%%%%%%%%%%%%%%%%%%%%%%%%%

\begin{footnotesize}

\end{footnotesize}

\end{document}